%% file: samplingfinal.tex
\documentclass[aps,pra,amsmath,amssymb,amsfonts,superscriptaddress]{revtex4}

\usepackage{epsfig,graphicx,bm,pstricks,pst-node,pst-text,pst-3d,amsthm}

\newcommand{\beq}{\begin{equation}}
\newcommand{\eeq}{\end{equation}}
\newcommand{\beqa}{\begin{eqnarray}}
\newcommand{\eeqa}{\end{eqnarray}}

\def\deq{\begin{equation}}
\def\feq{\end{equation}}
\def\deqn{\begin{eqnarray}}
\def\feqn{\end{eqnarray}}

\def\ketbra#1#2{|#2 \rangle  \langle #1| }
\def\undi{\mbox{1\hspace{-.15em}l}}

\newcommand{\ket} [1] {\vert #1 \rangle}
\newcommand{\bra} [1] {\langle #1 \vert}

\newcommand{\proj}[1]{\ket{#1}\bra{#1}}

\newtheorem{lem}{Lemma}
\newtheorem{thm}[lem]{Theorem}
\newtheorem{defn}[lem]{Definition}
\newtheorem{cor}[lem]{Corollary}

\newcommand{\abs}[1]{\left\vert#1\right\vert}

\newcommand{\protocol}[2]
{\begin{center}\framebox[1.1\width]{\begin{minipage}[l]{13cm}\begin{center}\underline{#1}\end{center}#2\end{minipage}}\end{center}}

\begin{document}

\title{Simulating quantum correlations as a distributed sampling problem}

\author{Julien Degorre}
\affiliation{Laboratoire de Recherche en Informatique, UMR 8263, Universit\'e Paris-Sud, 91405 Orsay, France}
\affiliation{Laboratoire d'Informatique Th\'eorique et Quantique,
D\'epartement d'Informatique et de Recherche Op\'erationnelle, Universit\'e de Montr\'eal, Canada}
\author{Sophie Laplante}
\author{J\'er\'emie Roland}
\affiliation{Laboratoire de Recherche en Informatique, UMR 8263, Universit\'e Paris-Sud, 91405 Orsay, France}

\begin{abstract}
It is known that quantum correlations exhibited by a maximally entangled qubit pair
can be simulated with the help of shared randomness,
supplemented with additional resources, such as communication,
post-selection or non-local boxes. For instance, in the case of projective measurements,
it is possible to solve this problem with protocols using
one bit of communication or making one use of a non-local box. We show that this problem
reduces to a distributed sampling problem. We give a new method to obtain samples from a biased
distribution, starting with shared random variables following a uniform distribution, and use it to build distributed
sampling protocols. This approach allows us to derive, in a simpler and unified way, many existing protocols
for projective measurements, and extend them to positive operator value measurements.
Moreover, this approach naturally leads to a local hidden variable model for Werner states.

\end{abstract}

\maketitle


\section{Introduction}
At the advent of quantum mechanics, some physicists were puzzled by the strange properties of quantum systems, compared
to classical physics, such as randomness and non-locality.
Einstein, Podolsky and Rosen showed that when two parties, say Alice and Bob, share an entangled state, the
outcome of a measurement on Alice's side is not only probabilistic, but may also be conditioned on the outcome of
a distant measurement on Bob's side.
Therefore, they questioned whether ``the quantum-mechanical description of physical reality could be considered complete''~\cite{epr35}.

To resolve this paradox, later called the EPR paradox, it was argued
that the apparent randomness in quantum experiments could actually come from
unknown ``hidden'' variables created locally along with the
supposedly quantum state, and that this randomness would disappear as soon as these hidden variables were revealed.
However, John Bell showed in 1964 that the quantum
correlations exhibited by the EPR \textit{gedanken} experiment, as reexpressed by Bohm~\cite{ba57},
could not be reproduced by so-called local hidden variable models, that is, models where Alice and Bob share an infinite
amount of locally created hidden variables~\cite{bell64}.
This proved in a sense the non-local character of quantum mechanics.

Recently, in order to understand quantum non-locality, with the help of the framework of communication complexity,
another approach has emerged to gauge the non-locality of quantum mechanics with
respect to local variable models. We know that if Alice and Bob share only a set
of local hidden variables (shared randomness), they cannot reproduce  quantum correlations, but if they
are allowed to use an additional resource, it may be possible for them to reproduce the
quantum correlations. The amount of additional resources used allows us to measure 
non-locality.

The most obvious resource that Alice and Bob can use in addition to shared
randomness is classical communication. The problem of simulating the quantum correlations with classical
communication was studied as early as 1992 by Maudlin~\cite{maudlin92}, who showed that
a finite amount of communication was enough, at least on average.
More precisely, in the case of projective measurements in the real plane on a maximally entangled qubit pair,
he gave a protocol using $1.17$ bits of communication on average,
but unbounded communication in the worst case.
This problem was revisited by Brassard \textit{et al} in 1999~\cite{bct99},
who showed that $8$ bits were sufficient in the worst case for arbitrary projective measurements.
These protocols use an infinite amount of shared randomness, and indeed
Massar \textit{et al.}~\cite{mbcc00} proved that the communication complexity
can be bounded in worst case only if the amount the shared randomness is infinite.
In 2000, inspired by Feldmann~\cite{feldmann95}, Steiner, independently of Maudlin,
showed that for projective measurements in the real plane,
$1.48$ bits were enough on average~\cite{steiner00}.
Cerf, Gisin and Massar~\cite{cgm00} proved that for an arbitrary projective measurement
$1.19$ bits of communication sufficed on average.
Finally in 2003, Toner and Bacon
\cite{tb03} showed that a local hidden variable model supplemented with one bit of
communication in the worst case is enough to reproduce the quantum correlations of the singlet
for arbitrary projective measurements.

Another resource which Alice and Bob can use to reproduce the quantum correlations
is post-selection. Here, Alice and Bob are allowed
to produce a special outcome of their measurement, noted $\bot$, meaning ``no result''.
This corresponds to the physical situation where
Alice and Bob's detectors are partially inefficient and sometimes do not click.
In 1999, Gisin and Gisin, inspired by Steiner's
communication protocol, gave a protocol which simulates quantum
correlations with shared randomness and with a probability $1/3$ of aborting
for either party.

Finally, Cerf \textit{et al.}~\cite{cgmp05} have shown that a third resource
could be used to simulate the quantum correlations: a non-local box.
The non-local box is a primitive shared between Alice and Bob, with two inputs and two
outputs, where the outputs  (conditioned on the inputs) are maximally non-local in the sense that they violate a Bell
inequality (CHSH) maximally while remaining causal. Cerf \textit{et al.}~\cite{cgmp05} have shown that
only one use of a non-local box suffices to simulate quantum correlations.

In this paper, we show that the
the problem can be reduced to a distributed sampling problem. In the local hidden
variable model for two parties, Alice has an input $a$ and an output $A$, and similarly,
Bob has an input $b$ and an output $B$, and they share a set of random variables
which are distributed independently
of Alice and Bob's input. Following an idea  introduced by Feldmann~\cite{feldmann95},
we can relax the condition that the shared randomness is
distributed independently of the input, and imagine that Alice and Bob share a set of
random variables with a distribution depending on Alice's input. Clearly in this scenario,
there exists a distribution which allows them to reproduce quantum correlations (a trivial
way is to let the random source produce $a$ with probability 1).
So the problem of reproducing the quantum correlations with different resources can be reduced
to the problem of Alice and Bob agreeing on a  sample from a distribution
depending on Alice's input.  We propose a method to carry out this distributed sampling
in two steps.  
The first is the completely local problem of how Alice can sample a
biased distribution depending on her input with the help of a (shared) uniform random source, and
the second step is how Alice can share this biased sample with Bob with the  help of various
resources. After giving a new method to perform the local
sampling, we will see that the second problem becomes easy, and allows us to understand  how
the various resources come into play. We reformulate previous protocols within this framework
in an intuitive and coherent way, including the best protocols using
communication, post-selection and non-local boxes to simulate the quantum correlations
for projective or POVM measurements, where we also extend previous results to protocols
using post-selection and non-local boxes.

The paper will be organized as follows: in Section~\ref{sec:correlations},
we will recall the EPR experiment and the LHV
model, and we extend it  to a setting where the parties share
a biased random source.
In Section~\ref{sec:local-sampling}, we will present two methods to perform the local sampling.
In Section~\ref{sec:distributed-sampling},
we will study the bipartite problem of how Alice can share her biased sample with
Bob. First, we will see the case where Alice and Bob do not use any resource, and see that
this results in a protocol  to simulate the projective
measurements on a Werner state. Second, we will study the case where Alice and Bob use different resources,
communication, post-selection, or non-local boxes in order
to simulate projective measurements on the singlet state.
Finally, we study the previous problem for generalized measures (POVM).


\section{Simulating the quantum correlations\label{sec:correlations}}
\subsection{Quantum correlations}
\begin{figure}[htb]
\begin{center}
\input{epr}
\end{center}
\caption{EPR experiment. Alice and Bob share a pair of qubits in the singlet state $\ket{\psi^-}=(\ket{\!\uparrow\downarrow}-\ket{\!\downarrow\uparrow})/\sqrt{2}$. Both perform a measurement
 on their qubit, specified by vectors $\vec{a}$ or $\vec{b}$, and obtain results $A=\pm 1$ or $B=\pm 1$.\label{fig:epr}}
\end{figure}
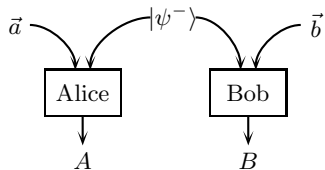

Let us recall Bohm's version~\cite{ba57} of the EPR \textit{gedanken} experiment (see Fig.~\ref{fig:epr}):
\begin{defn}[EPR Experiment]\label{def:epr}
Two parties, Alice and Bob, share a qubit pair in the singlet state,
$\ket{\psi^-}=($ $\ket{\!\uparrow\downarrow}$  $-\ket{\!\downarrow\uparrow})/{\sqrt{2}}$,
that is, a maximally entangled state of two qubits. Alice and Bob then each receive the classical description of a
projective measurement they have to perform on their respective qubit. These can be represented by unit vectors $\vec{a}$ and $\vec{b}$ pointing in some direction on the Bloch sphere. They then obtain measurement outcomes $A\in\{1,-1\}$ and $B\in\{1,-1\}$
respectively, where $+1$ corresponds to a spin parallel to the measurement direction, and $-1$ to a spin anti-parallel to this direction.
\end{defn}
According to quantum mechanics, the outcome of Alice's and Bob's measurements,
$A$ and $B$, have the following joint probabilities:
\deq \label{eq:probas}
p(A,B)=\frac{1-AB\ \vec{a}\cdot\vec{b}}{4},
\feq
or, equivalently, their joint and marginal expectation values are given by
\beqa
E(AB|\vec{a},\vec{b})&=&-\vec{a}\cdot\vec{b}, \label{eq:correlation}\\
E(A|\vec{a},\vec{b})&=&0,\\
E(B|\vec{a},\vec{b})&=&0.
\eeqa

\subsection{Local hidden variable models}
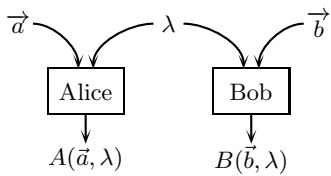
\begin{figure}[htb]
\begin{center}
\input{lhv}
\end{center}
\caption{LHV model for the EPR experiment.\label{fig:lhv}}
\end{figure}

As pointed out by Einstein, Podolsky and Rosen, the correlations between Alice's and Bob's measurement outcomes
in such a \textit{gedanken} experiment show a disturbing property of quantum mechanics, through a kind of influence at a distance
between Alice's and Bob's outcomes, this is the EPR paradox~\cite{epr35}.
To circumvent this paradox,
the following, completely classical, model was proposed to simulate the EPR experiment (see Fig.~\ref{fig:lhv})~\cite{bell64}:
\begin{defn}[Local hidden variable model]
Alice and Bob share some random variable $\lambda\in\Lambda$, where $\Lambda$ is some possibly
infinite set, with probability distribution $p(\lambda)$. They then receive inputs
$\vec{a}\in\mathbb{S}_2$ and $\vec{b}\in\mathbb{S}_2$ respectively, and output $A=A(\vec{a},\lambda)\in\{1,-1\}$ and
$B=B(\vec{b},\lambda)\in\{1,-1\}$ respectively.
\end{defn}
In this scenario the functions $A(\vec{a},\lambda)$ and $B(\vec{b},\lambda)$ are deterministic so that,
as intended, Alice's and Bob's outputs are fixed as soon as the value of the random variable $\lambda$ is known.
Moreover, this hidden variable may have been created locally (as was the quantum state in the EPR experiment)
and then communicated to both Alice and Bob.
Therefore, if such a model could reproduce the quantum correlations of the EPR experiment described above,
it could conveniently replace quantum mechanics as it would circumvent not only randomness but also non-locality.

However, Bell has shown in his famous theorem that such a simulation is not possible~\cite{bell64}.
Indeed, let us note that we have implicitly assumed that the hidden variable $\lambda$ was distributed independently of the measurement
directions $\vec{a}$ and $\vec{b}$, $p(\lambda|\vec{a},\vec{b})=p(\lambda)$,
since the physical interpretation is that $\lambda$ has been created along with the supposedly quantum state, which could have happened
long before the inputs $\vec{a}$ and $\vec{b}$ were fixed.
\begin{thm}[Bell]\label{thm:bell}
No local hidden variable model may simulate the quantum correlations exhibited by the EPR experiment
given in Definition~\ref{def:epr}.
\end{thm}

\begin{proof} (Sketch)
For a local hidden variable model, we have
$E(AB|\vec{a},\vec{b})=\sum_{\lambda\in\Lambda} p(\lambda)\ A(\vec{a},\lambda)B(\vec{b},\lambda)$.
It is then easy to show that all local hidden variable models satisfy the CHSH inequality~\cite{chsh69}
\beq
-2\leq C(\vec{a_1},\vec{a_2},\vec{b_1},\vec{b_2})\leq 2\quad (\forall \vec{a_1},\vec{b_1},\vec{a_2},\vec{b_2}\in\mathbb{S}_2),
\eeq
where
\beq
C(\vec{a_1},\vec{a_2},\vec{b_1},\vec{b_2})
=E(AB|\vec{a_1},\vec{b_1})+E(AB|\vec{a_1},\vec{b_2})+E(AB|\vec{a_2},\vec{b_1})-E(AB|\vec{a_2},\vec{b_2}).
\eeq
Nevertheless, for quantum mechanics, there exist $\vec{a_1},\vec{a_2},\vec{b_1},\vec{b_2}\in\mathbb{S}_2$ such that Eq.~(\ref{eq:correlation}) yields
$C(\vec{a_1},\vec{a_2},\vec{b_1},\vec{b_2})= 2\sqrt{2}>2$, so that quantum mechanics
violates the CHSH inequality and therefore cannot be reproduced by a local hidden variable model.
\end{proof}


\subsection{Protocol with a biased random source}
We have seen that a local hidden variable model, as defined above, where the random variable $\lambda$
was distributed independently of the inputs $\vec{a}$ and $\vec{b}$,
 does not allow for the simulation of the EPR experiment. Nonetheless, if we relax our model and let the random variable
depend on one of the inputs, it becomes possible to reproduce the quantum joint distribution (\ref{eq:probas})
(this is a slight extension of a result by Feldmann~\cite{feldmann95}):
\begin{thm}[Sampling theorem]\label{thm:sampling}
Let $\vec{a}$ and $\vec{b}$ be Alice's and Bob's inputs. If Alice and Bob share a random variable $\vec{\lambda_s}\in\mathbb{S}_2$ distributed according to a
biased distribution with probability density \beq
\rho(\vec{\lambda_s}|\vec{a},\vec{b})=\rho_{\vec{a}}(\vec{\lambda_s})=\frac{\abs{\vec{a}\cdot\vec{\lambda_s}}}{2\pi}, \eeq then they are able to simulate the
EPR experiment in Definition~\ref{def:epr} without any further resource, that is, simulating the
EPR experiment reduces to distributed sampling from $\rho_{\vec{a}}$.
\end{thm}
\begin{proof}
If Alice and Bob set their respective outputs as
$A(\vec{a},\vec{\lambda_s})=-\textrm{sgn}(\vec{a}\cdot\vec{\lambda_s})$ and $B(\vec{b},\vec{\lambda_s})=\textrm{sgn}(\vec{b}\cdot\vec{\lambda_s})$,
where $\textrm{sgn}(x)=1$ for $x\geq 0$ and $\textrm{sgn}(x)=-1$ for $x<0$ $(x\in\mathbb{R})$,
then the joint expectation value $E(AB|\vec{a},\vec{b})$ is given by
\beqa
E(AB|\vec{a},\vec{b})&=&\int_{S_2} \rho(\vec{\lambda_s}|\vec{a},\vec{b})\ A(\vec{a},\vec{\lambda_s})B(\vec{b},\vec{\lambda_s})\ d\vec{\lambda_s}\\
&=&-\frac{1}{2\pi}\int_{S_2} (\vec{a}\cdot\vec{\lambda_s})\ \textrm{sgn}(\vec{b}\cdot\vec{\lambda_s})\ d\vec{\lambda_s}\\
&=&-\vec{a}\cdot\vec{b}.
\eeqa
Similarly, we have $E(A|\vec{a},\vec{b})=E(B|\vec{a},\vec{b})=0$, as desired.
\end{proof}

Theorem~\ref{thm:sampling} allows us to reduce the simulation of quantum correlations to the distributed
sampling of a biased distribution $\rho_{\vec{a}}$.
The problem can further be reduced to two steps. First, Alice can locally create a sample $\vec{\lambda_s}\sim\rho_{\vec{a}}$ using her knowledge of
$\vec{a}$ and uniformly distributed random variables. This is what we will call local sampling.
The second step is a communication complexity problem: Alice shares this biased variable with  Bob.
For this step, they will need resources in addition to those allowed by local hidden variable models.
Let us now study the first problem, the local sampling.


\section{Local sampling of the biased distribution\label{sec:local-sampling}}
\subsection{The rejection method}
From now on, we will use the notation $\vec{\lambda_s}$ for the biased samples, $(\vec{\lambda_0},\vec{\lambda_1},\ldots)$
for the sequence of uniformly distributed random variables shared by Alice and Bob, and $\sqcup_\Lambda$ for a uniform distribution
on the set $\Lambda$.

Forgetting about Bob, the problem is for Alice to sample from the biased
distribution $\rho_{\vec{a}}$ from a source of uniformly distributed variables.
A well known method to perform (local) sampling is the rejection method~\cite{devroye}, which in our case gives:
\begin{thm}[The rejection method]\label{thm:rejection}
Let Alice perform the following protocol:
\protocol{Rejection method}{
Set $k=0$
\begin{enumerate}
\item Alice picks $\vec{\lambda_k}\sim\sqcup_{\mathbb{S}_{2}}$,
\item Alice picks $u_k\sim \sqcup_{[0,1]}$,
\item If $u_k\leq|\vec{a}\cdot\vec{\lambda_k}|$,
then she accepts $\vec{\lambda_k}$ and sets $\vec{\lambda_s}=\vec{\lambda_k}$,\\
otherwise, she rejects $\vec{\lambda_k}$ and goes back to step $1$ with $k=k+1$.
\end{enumerate}}
When the process terminates, we have $\vec{\lambda_s}\sim{|\vec{a}\cdot\vec{\lambda_s}|}/{2\pi}$.
\end{thm}

Let us note that this method is iterative: if Alice rejects $\vec{\lambda_k}$, she has to start over with a fresh random variable $\vec{\lambda}_{k{+}1}$, and
so on until she accepts one sample. If Alice is particularly unlucky, she could reject an arbitrarily large number of samples before accepting one. This means
that in the worst case the process takes unbounded time to terminate and requires an infinite amount of samples.

\subsection{The ``choice'' method}
To avoid this drawback, we now propose a new method, that we will call the ``choice'' method.
Contrary to the rejection method, this method will not reject anything, but it is less general
than the rejection method because it makes use of a specific property of the biased distribution that we want to sample from.
Indeed, the key is the following remark: in the rejection method above, the bias
$|\vec{a}\cdot\vec{\lambda_0}|$ is uniformly distributed in $[0,1]$ when
$\vec{\lambda_0} \sim \sqcup_{\mathbb{S}_{2}}$. Therefore, we could produce a sample $u_0\sim\sqcup_{[0,1]} $,
by picking a second vector uniformly distributed on the sphere $\vec{\lambda_1}\sim\sqcup_{\mathbb{S}_{2}}$, and setting
$u_0=|\vec{a}\cdot\vec{\lambda_1}|$.
Hence, to sample from the biased distribution $\rho_{\vec{a}}(\vec{\lambda_s})={|\vec{a}\cdot\vec{\lambda_s}|}/{2\pi}$,
we may use the following theorem:
\begin{thm}[The ``choice'' method]\label{thm:choice}
If Alice performs the following protocol:
\protocol{Choice method}{
\begin{enumerate}
\item Alice picks  $\vec{\lambda_0} \sim \sqcup_{\mathbb{S}_{2}}$,
\item Alice picks  $\vec{\lambda_1} \sim \sqcup_{\mathbb{S}_{2}}$,
\item If $|\vec{a}\cdot\vec{\lambda_1}|\leq|\vec{a}\cdot\vec{\lambda_0}|$,
then she accepts $\vec{\lambda_0}$ and sets $\vec{\lambda_s} = \vec{\lambda_0}$, \\
otherwise, she accepts $\vec{\lambda_1}$ and  sets $\vec{\lambda_s} = \vec{\lambda_1}$.
\end{enumerate}}
then $\vec{\lambda_s} \sim {|\vec{a}\cdot\vec{\lambda_s}|}/{2\pi}$
and $p(\vec{\lambda_s}=\vec{\lambda_0})=p(\vec{\lambda_s}=\vec{\lambda_1})=\frac{1}{2}$.
\end{thm}
\begin{proof}
Let  $u_0=|\vec{a}\cdot\vec{\lambda_1}|$ and $u_1=|\vec{a}\cdot\vec{\lambda_0}|$.
Then $u_0,u_1\sim\sqcup_{[0,1]}$ when $\vec{\lambda_0},\vec{\lambda_1}\sim\sqcup_{\mathbb{S}_{2}}$,
so the rejection method ensures that conditionally to the test $u_0\leq|\vec{a}\cdot\vec{\lambda_0}|$ being satisfied,
$\vec{\lambda_s}$ will be distributed according to the distribution $\rho_{\vec{a}}$.
Exchanging $\vec{\lambda_0}$ and $\vec{\lambda_1}$'s roles, the same argument ensures that conditionally to the
test $u_1<|\vec{a}\cdot\vec{\lambda_1}|$ being satisfied,
$\vec{\lambda_s}$ will also be distributed according to $\rho_{\vec{a}}$, so we have
$
\rho(\vec{\lambda_s}|u_0\leq|\vec{a}\cdot\vec{\lambda_0}|)=
\rho(\vec{\lambda_s}|u_1<|\vec{a}\cdot\vec{\lambda_1}|)=\rho_{\vec{a}}(\vec{\lambda_s})
$.
As the outcomes of the test are of course mutually
exclusive, $\vec{\lambda_s}$ will always have the right distribution.
\end{proof}


\section{Distributed sampling and simulating correlations\label{sec:distributed-sampling}}
\subsection{Simulation of a Werner state}
A Werner state is a quantum state obtained by mixing a pure state $\ket{\psi}$ with maximally random noise. It is characterized
by a density matrix $W=p\proj{\psi}+(1-p)\undi/d$, where $d$ is the dimension of the Hilbert space in which $\ket{\psi}$ lies,
and $p$ is the so-called visibility of $\ket{\psi}$ in the Werner state.
These states were introduced by Werner to show
that the correlations exhibited by some non-separable states could be simulated by local hidden variable models~\cite{werner89}.
We will now show that our approach leads naturally to a local hidden variable model for a Werner state of the singlet $\ket{\psi^-}$.

In the previous section, we have given two methods for Alice to sample from the specific biased distribution
using uniformly distributed variables. Suppose Alice performs the choice method. Without any further resource
than that allowed by local hidden variable models, Bob is not aware of Alice's choice, and hence does not share
with Alice the variable $\vec{\lambda_s}\sim \rho_{\vec{a}}$. Because of Bell's theorem, we know
that Alice and Bob will not be able to simulate the singlet correlations, but Bob could nevertheless
try to guess $\vec{\lambda_s}$. More specifically, if he assumes that
$\vec{\lambda_s}=\vec{\lambda}_0$, he will be right half of the time. As an intermediate step
towards simulating singlet correlations, we show that with such
a local hidden variable model, Alice and Bob may simulate a Werner state of the singlet state
with a visibility of one half.

\begin{thm}[Simulation of Werner states]
A local hidden variable model may simulate the correlations exhibited by the Werner state
\beq\label{thm:werner}
W=p\ \ketbra{\psi^-}{\psi^-}+(1-p)\ \frac{\undi}{4}
\eeq
with a visibility of $p=1/2$.
\end{thm}
\begin{proof}
As already stated, the local hidden variable model consists for Alice to perform the choice method, and for Bob to always assume that
Alice chose $\vec{\lambda_0}$:
\protocol{LHV model for Werner state}{
\begin{enumerate}
\item Alice and Bob share a pair of uniformly distributed variables $\vec{\lambda_0},\vec{\lambda_1}\sim\sqcup_{\mathbb{S}_{2}}$,
\item Alice performs the choice method: she tests whether $|\vec{a}\cdot\vec{\lambda_1}|\leq|\vec{a}\cdot\vec{\lambda_0}|$,
\begin{itemize}
\item if so, she outputs $A=-\textrm{sgn}(\vec{a}\cdot\vec{\lambda_0})$,
\item otherwise, she outputs $A=-\textrm{sgn}(\vec{a}\cdot\vec{\lambda_1})$,
\end{itemize}
\item Bob outputs $B=\textrm{sgn}(\vec{b}\cdot\vec{\lambda_0})$.
\end{enumerate}}

To prove that this model achieves its goal, we first note that Alice's output may be rewritten in the more compact
form $A=-\textrm{sgn}(\vec{a}\cdot(\vec{\lambda_0}+\vec{\lambda_1}))$. The local hidden variable then yields:
\beqa
E(AB|\vec{a},\vec{b})&=&-\frac{1}{(4\pi)^2}\int_{\mathbb{S}_{2}}d\vec{\lambda_0}\int_{\mathbb{S}_{2}}d\vec{\lambda_1}\
\textrm{sgn}(\vec{a}\cdot(\vec{\lambda_0}+\vec{\lambda_1}))\ \textrm{sgn}(\vec{b}\cdot\vec{\lambda_0})\\
&=&-\frac{\vec{a}\cdot\vec{b}}{2},
\eeqa
that is the same joint expectation value as for the Werner state (\ref{thm:werner}) with a visibility of $p=1/2$.
As we also have $E(A|\vec{a},\vec{b})=E(B|\vec{a},\vec{b})=0$, this model reproduces all the correlations exhibited by this state.
\end{proof}

Actually, even though we have derived our local hidden variable in a quite different way, it finally reduces to a rewriting
of the model proposed by Werner in~\cite{werner89}.

\subsection{Simulation of the singlet state with additional resources between Alice and Bob}
Now, if Alice and Bob want to simulate the correlations of the singlet with perfect visibility, we know from Bell's
theorem (Theorem~\ref{thm:bell}), that they will need additional resources.
Moreover, we know that they may achieve their goal as soon as they share some random variable $\vec{\lambda_s}$
distributed according to the biased distribution $\rho_{\vec{a}}$.
We will now consider three different resources that will allow Alice and Bob to sample from this distribution:
post-selection, classical communication, and non-local boxes. Let us note that in this section, we will not introduce
any new protocols, but rather we will show that the most efficient known protocols can be viewed as
sampling from the same biased distribution $\rho_{\vec{a}}$. This reduction to the distributed sampling problem
is our main contribution.

\subsubsection{Post-selection}
Let us first consider post-selection, which is the possibility for Alice and Bob to sometimes abort
the protocol. Physically this would correspond to the fact that Alice and Bob's detectors are imperfect
and sometimes do not click. In our model, post-selection is represented by a new symbol $\bot$ added
to the set of valid outputs for Alice and Bob.
We will use the notation $\bot_A$ for the event that Alice aborts, and similarly $\bot_B$ for Bob.

\begin{thm}[Post-selection]\label{thm:post-selection}
If Alice and Bob are given an infinite amount of shared randomness, supplemented with post-selection
for only one of them, say Alice, there exists a protocol with $p(\bot_A)=1/2$
such that conditionally to the fact that no party aborted, Alice and Bob share
a random variable $\vec{\lambda_s}\sim \rho_{\vec{a}}$.
\end{thm}
Together with Theorem~\ref{thm:sampling}, this implies the following corollary
\begin{cor}\label{cor:gisin-gisin}
Using a local hidden variable model supplemented with
post-selection, it is possible for Alice and Bob to simulate the EPR experiment in Definition~\ref{def:epr}
with $p(\bot_A)=1/2$ and $p(\bot_B)=0$.
\end{cor}

\begin{proof}[Proof of Theorem~\ref{thm:post-selection}]
The idea is to let Alice perform the rejection method
(Theorem~\ref{thm:rejection}) and use post-selection to reject the bad $\vec{\lambda}$'s, which leads to the following
protocol:
\protocol{Post-selection protocol}
{\begin{enumerate}
\item Alice and Bob share a uniformly distributed variable $\vec{\lambda}\sim\sqcup_{\mathbb{S}_{2}}$,
\item Alice picks $u\sim\sqcup_{[0,1]}$,
\item She performs the rejection method: she tests whether $u\leq|\vec{a}\cdot\vec{\lambda}|$,
  \begin{itemize}
  \item if so, she sets $\vec{\lambda_s}=\vec{\lambda}$,
  \item otherwise, she \textit{aborts the protocol} ($\bot_A$),
  \end{itemize}
\item Bob always sets $\vec{\lambda_s}=\vec{\lambda}$.
\end{enumerate}
}
Since both $u$ and $|\vec{a}\cdot\vec{\lambda}|$ are uniformly distributed on $[0,1]$ when $\vec{\lambda}\sim\sqcup_{\mathbb{S}_{2}}$,
it is clear that $p(\bot_A)=p(u\leq|\vec{a}\cdot\vec{\lambda}|)=1/2$. Moreover, Theorem~\ref{thm:rejection} ensures that
$\rho(\vec{\lambda_s}|\neg\bot_A)=\rho_{\vec{a}}(\vec{\lambda_s})$.
\end{proof}


Let us note that this is just a rewriting of Gisin and Gisin's protocol~\cite{gg99}. Here, only Alice uses post-selection, but, as pointed out by these authors, it is clearly possible to symmetrize the protocol
by randomly exchanging Alice's and Bob's role, and using an
additional shared random variable to tell Alice and Bob to both abort simultaneously
(see~\cite{gg99} for details). This leads to a protocol with $p(\bot_A)=p(\bot_B)=1/3$.

\subsubsection{Communication}
Let us now consider a second resource used to sample from $\rho_{\vec{a}}$: communication between Alice and Bob.
First of all, we briefly recall Steiner's protocol~~\cite{steiner00}, which, in a similar manner to Gisin and Gisin's, consists in
using the rejection method to sample from the biased distribution.
His idea was to consider that Alice and Bob share an infinite sequence
$(\vec{\lambda_0},\vec{\lambda_1},\cdots)$ of samples $\vec{\lambda_k}\sim\sqcup_{\mathbb{S}_{2}}$
(recall that the local hidden variable model assumes no limitation on the amount of shared randomness).
Similarly to Gisin and Gisin's protocol, Alice then performs the rejection method with the first sample
$\vec{\lambda_0}$, but instead of aborting the protocol if she has to reject $\vec{\lambda_0}$,
she iterates the method by taking the samples in the order of the sequence until she accepts
one of them, say $\vec{\lambda_k}$. She then communicates the index $k$ of the accepted sample to Bob.
As pointed out earlier, if Alice is particularly unlucky, she could reject an
arbitrarily large number of $\vec{\lambda}$'s before accepting one. The index $k$ she has to communicate
to Bob may then become arbitrarily large, such that in the worst case the amount of communication required
to simulate the EPR experiment with this method is unbounded.

Instead of using the rejection method, which leads to unbounded communication complexity
in the worst case, we can use the choice method to obtain a protocol with bounded
communication in the worst case.
\begin{thm}[Communication]\label{thm:communication}
If Alice and Bob are given an infinite amount of shared randomness, supplemented with one-way
communication, there exists a protocol using exactly one bit of communication
that allows Alice and Bob to share a random variable $\vec{\lambda_s}\sim \rho_{\vec{a}}$.
\end{thm}
Together with Theorem~\ref{thm:sampling}, we have the following corollary:
\begin{cor}\label{cor:toner-bacon}
Using a local hidden variable model supplemented with
one-way communication, it is possible for Alice and Bob to simulate the EPR experiment in Definition~\ref{def:epr}
with exactly one bit of communication.
\end{cor}

\begin{proof}[Proof of Theorem~\ref{thm:communication}]
Let the two parties use the following protocol, where Alice performs the choice method
shown in Theorem~\ref{thm:choice}, and then communicates to Bob the index of the sample she accepted:
\protocol{Communication protocol}{
\begin{enumerate}
\item Alice and Bob share a pair of uniformly distributed variables $\vec{\lambda_0},\vec{\lambda_1} \sim \sqcup_{\mathbb{S}_{2}}$,
\item Alice performs the choice method: she tests whether $|\vec{a}\cdot\vec{\lambda_1}|\leq|\vec{a}\cdot\vec{\lambda_0}|$,
  \begin{itemize}
  \item if so, she accepts $\vec{\lambda_0}$ and sets $x=0$,
  \item otherwise, she accepts $\vec{\lambda_1}$ and sets $x=1$,
  \end{itemize}
\item Alice sends $x$ to Bob,
\item Alice and Bob then set $\vec{\lambda_s}=\vec{\lambda_x}$.
\end{enumerate}}
Hence, Theorem~\ref{thm:choice} ensures that $\vec{\lambda_s} \sim {|\vec{a}\cdot\vec{\lambda_s}|}/{2\pi}$.
\end{proof}

Let us note that Corollary~\ref{cor:toner-bacon} was proven by Toner and Bacon~\cite{tb03}, but our approach has the advantage
of giving much more intuition, and in particular of clarifying the relationship with the previous protocols.

\subsubsection{Non-local boxes}
We now consider another additional resource shared by Alice and Bob: a non-local box. This resource was introduced by Popescu and Rohrlich~\cite{pr97}, and has recently been used by Cerf, Gisin, Massar and Popescu
to simulate the singlet correlations~\cite{cgmp05}.
Let us recall its definition and its main characteristics.
\begin{defn}
A PR non-local box is a device shared by Alice and Bob, that has two input bits
$x,y \in \{0,1\}$ from Alice and Bob, respectively, and outputs $\alpha,\beta \in \{0,1\}$ to Alice and Bob,
respectively, according to the following distribution:
\beq
p(\alpha,\beta|x,y)=\left\{\begin{array}{cc}
\frac{1}{2} & \textrm{if}\ \alpha\oplus \beta=x\wedge y,\\
0 & \textrm{otherwise.}
\end{array}\right.
\eeq
\end{defn}
One use of a non-local box will be called an nl-bit.
This resource has the following interesting properties:
\begin{itemize}
\item it is maximally non-local, in the sense that it maximally violates the CHSH Bell inequality.
\item it is causal, in the sense that Alice's output $\alpha$
is independent of Bob's input $y$, $p(\alpha|x,y)=p(\alpha|x)$ (and vice versa),
\item it is a strictly weaker resource than one bit of communication: due to the causality property,
it may not be used to communicate but, on the other hand, it may be shown that one use of a non-local
box may be simulated by one bit of communication~\cite{cgmp05}.
\end{itemize}

Since we have shown that the previous protocols for simulating the singlet correlations
reduced to sampling from the distribution $\rho_{\vec{a}}$, and
motivated by Cerf, Gisin, Massar and Popescu's result~\cite{cgmp05}, we could be tempted to study
whether a non-local box also allows Alice and Bob to share a random variable $\vec{\lambda_s}\sim\rho_{\vec{a}}$.
However, this is obviously not possible since $\vec{\lambda_s}$ is not independent from $\vec{a}$, so that
their mutual information is non-zero, $I(\vec{\lambda_s}:\vec{a})\neq 0$. Indeed, if Bob knew $\vec{\lambda_s}$,
this would mean that he has gained some information about $\vec{a}$, but this is impossible using only
shared randomness and a non-local box since we have seen that a non-local box does not allow signaling.

However, by examining the proof of Theorem~\ref{thm:sampling}, we notice that to simulate the singlet correlations from
a shared random variable $\vec{\lambda_s}\sim\rho_{\vec{a}}$, Bob actually only uses the knowledge
of 
$\textrm{sgn}(\vec{b}\cdot\vec{\lambda_s})$,
(and similarly for Alice, but this is not really an issue). Let us introduce the following definition:
\begin{defn}\label{def:f-sampling}
Let $\vec{a},\vec{b}$ be Alice's and Bob's inputs. A protocol is an $f$-sampling protocol for a given distribution $\rho$
if at the end, Alice has a sample $s_A=f(\vec{a},\vec{\lambda_s})$ and Bob has a sample
$s_B=f(\vec{b},\vec{\lambda_s})$, where $\vec{\lambda_s}\sim\rho$.
\end{defn}

We may now rewrite Theorem~\ref{thm:sampling} with a weaker sampling hypothesis:
\begin{thm}[Strong sampling theorem]\label{thm:strong-sampling}
The simulation of the EPR experiment in Definition~\ref{def:epr} is equivalent to
$f$-sampling
for $\rho_{\vec{a}}$, where
$\rho_{\vec{a}}(\vec{\lambda_s})={|{\vec{a}\cdot\vec{\lambda_s}}|}/{2\pi}$, and
$f(\vec{x},\vec{y})=\textrm{sgn}(\vec{x}\cdot\vec{y})$.
\end{thm}
\begin{proof}
The proof  that simulating the EPR experiment reduces to $f$-sampling
for $\rho_{\vec{a}}$ is
identical to Theorem~\ref{thm:sampling}.
To show that $f$-sampling
for $\rho_{\vec{a}}$ reduces to simulating the EPR experiment,
let $A,B$ be the outcome of a simulation of the EPR experiment.
Let $s_A=-A$ and $s_B=B$.  Then  $s_A,s_B$ is an $f$-sample
for $\rho_{\vec{a}}$.

\end{proof}

With this weaker hypothesis in mind, we will now prove the following theorem:
\begin{thm}[Non-local box]\label{thm:nl-box}
If Alice and Bob are given an infinite amount of shared randomness,
there exists an $f$-sampling protocol for $\rho_{\vec{a}}$ with $f(\vec{x},\vec{y})=\textrm{sgn}(\vec{x}\cdot\vec{y})$
that only makes use of $1$ nl-bit.
\end{thm}
Once again, together with Theorem~\ref{thm:strong-sampling}, we have the following corollary:
\begin{cor}\label{cor:cgmp}
Using a local hidden variable model supplemented with a shared PR non-local box,
it is possible for Alice and Bob to simulate the EPR experiment in Definition~\ref{def:epr}
with exactly $1$ nl-bit.
\end{cor}

\begin{proof}[Proof of Theorem~\ref{thm:nl-box}]
Before introducing the protocol, let us give some motivation by recalling the protocol that
simulates the Werner state with visibility $p=1/2$ (see Theorem~\ref{thm:werner}).
In this protocol, Alice performs the choice method (she chooses between
$\vec{\lambda_0}$ and $\vec{\lambda_1}$) and Bob always assumes that she chose $\vec{\lambda_0}$.
Obviously, he will be wrong half of the times, when Alice chose $\vec{\lambda_1}$, but sometimes
this is not a problem since it may happen that $\textrm{sgn}(\vec{b}\cdot\vec{\lambda_0})=\textrm{sgn}(\vec{b}\cdot\vec{\lambda_1})$.
The idea of the following protocol is to use a PR non-local box to correct Bob's remaining mistakes, which
happens when Alice chooses $\vec{\lambda_1}$ and $\textrm{sgn}(\vec{b}\cdot\vec{\lambda_0})\neq \textrm{sgn}(\vec{b}\cdot\vec{\lambda_1})$.

\protocol{Non-local box protocol}{
\begin{enumerate}
\item Alice and Bob share a pair of uniformly distributed variables $\vec{\lambda_0},\vec{\lambda_1} \sim \sqcup_{\mathbb{S}_{2}}$,
\item Alice performs the choice method: she tests whether $|\vec{a}\cdot\vec{\lambda_1}|\leq|\vec{a}\cdot\vec{\lambda_0}|$,
  \begin{itemize}
  \item if so, she accepts $\vec{\lambda_0}$ and sets $x=0$,
  \item otherwise, she accepts $\vec{\lambda_1}$ and sets $x=1$,
  \end{itemize}
\item Alice inputs $x$ into the non-local box,
\item Bob tests whether $\textrm{sgn}(\vec{b},\vec{\lambda_0})=\textrm{sgn}(\vec{b},\vec{\lambda_1})$
  \begin{itemize}
  \item if so, he sets $y=0$,
  \item otherwise, he sets $y=1$,
  \end{itemize}
\item Bob inputs bit $y$ into the non-local box,
\item Alice gets output bit $\alpha$ from the box and sets $\vec{\lambda_s}=(-1)^\alpha\vec{\lambda_x}$,
\item Bob gets output bit $\beta$ from the box and sets $\vec{\lambda_B}=(-1)^\beta\vec{\lambda_0}$.
\end{enumerate}}

Let us now analyse the protocol. As Alice uses the choice method to choose $\vec{\lambda_x}$, we know from Theorem~\ref{thm:choice}
that $\vec{\lambda_x}\sim\rho_{\vec{a}}$. Moreover, since $\alpha$ is an unbiased random bit,
$\vec{\lambda_s}=(-1)^\alpha\vec{\lambda_x}$ is also distributed according to $\rho_{\vec{a}}$.
It is clear that Alice may compute
$s_A=\textrm{sgn}(\vec{a}\cdot\vec{\lambda_s})$ since she knows $\vec{a}$ and $\vec{\lambda_s}$.

On Bob's side, an error takes place if and only if  $\vec{\lambda_x}=\vec{\lambda_1}\ \textrm{and}\ \textrm{sgn}(\vec{b},\vec{\lambda_0})\neq\textrm{sgn}(\vec{b},\vec{\lambda_1})$, and this
is precisely the case when the non-local box's outputs are different:
\beq
[\vec{\lambda_x}=\vec{\lambda_1}\ \textrm{and}\ \textrm{sgn}(\vec{b},\vec{\lambda_0})\neq\textrm{sgn}(\vec{b},\vec{\lambda_1})]
    \Longleftrightarrow
[x\wedge y]
    \Longleftrightarrow
\beta\neq\alpha.
\eeq
Therefore, the non-local box corrects Bob's error by changing  the sign of $\vec{\lambda_B}=(-1)^\beta\vec{\lambda_0}$ with respect to $\vec{\lambda_s}$ when necessary,
so that he may correctly compute $s_B=\textrm{sgn}(\vec{b}\cdot\vec{\lambda}_B)=\textrm{sgn}(\vec{b}\cdot\vec{\lambda_s})$.
\end{proof}

Let us note that this reproves Cerf, Gisin, Massar and Popescu's result~\cite{cgmp05}, but with a more intuitive approach.

\subsection{Simulation of POVMs with additional resources}
In the last section, we derived protocols for simulating local projective measurements on the singlet.
However, quantum mechanics allows a wider class of
measurements than projective measurements, and we will now use our approach to simulate the EPR experiment on the singlet with the most general type
of measurement allowed by quantum mechanics, namely positive operator value measure (POVM).
Let us first recall some definitions.
\begin{defn}[POVM]\label{def:povm}
A positive operator value measure $\mathcal{A}$ of size $r$ on a Hilbert space
$\mathcal{H}_n$ of dimension $n$ is a collection of $r$ positive (i.e. with non-negative eigenvalues)
operators on $\mathcal{H}_n$, $\{A_i|i=1,\ldots,r\}$, such that
\beq\label{eq:povm}
\sum_{i=0}^{r} A_i = \undi_n,
\eeq
where $\undi_n$ is the identity on $\mathcal{H}_n$.
\end{defn}

The projective measurements that we have considered above are a particular case of POVM where the
elements $A_i$ are orthogonal rank one projectors, that is $A_i A_j = \delta_{ij} A_i$.
In other words, they define a basis $\{\ket{\phi_i}|i=1,\ldots,r\}$ such that $A_i=\proj{\phi_i}$.

Without loss of generality, we can restrict our study to rank one POVMs,
i.e.,~POVMs where the elements $A_i$ are proportional to rank one projectors
$\proj{\phi_i}$.
In dimension $2$, that is for a qubit, a rank one projector $\proj{\phi_i}$ may be represented
on the Bloch sphere by
a unit vector $\vec{v}$ such that $\proj{\phi_i}=(\undi_2+\vec{v}\cdot\vec{\sigma})/2$, where
$\vec{\sigma}$ is the vector of Pauli matrices.
Therefore, the elements $A_i$ of a rank one POVM may be represented
by a (not necessarily unit) vector $\vec{a_i}$ such that
$A_i=(|\vec{a_i}|\undi_2+\vec{a}_i\cdot\vec{\sigma})/2$.

From Definition~\ref{def:povm}, we see that a rank one POVM $\mathcal{A}$
of size $r$ on a qubit may be represented by a set of $r$ vectors in the Bloch ball,
$\{\vec{a_i}|i=1,\ldots,r\}$ satisfying the following conditions, equivalent to (\ref{eq:povm}):
\beqa
\sum_{i=0}^{r}|\vec{a_i}|&=&2\label{eq:cond-povm1},\\
\sum_{i=0}^{r}\vec{a_i}&=&0\label{eq:cond-povm2}.
\eeqa

Now we generalize the EPR experiment in Definition~\ref{def:epr} to the case where Alice and Bob perform POVMs.
\begin{defn}[EPR Experiment with POVMs]\label{def:eprPovm}
Two parties, Alice and Bob, share a qubit pair in the singlet state,
$\ket{\psi^-}=(\ket{\uparrow\downarrow}-\ket{\downarrow\uparrow})/{\sqrt{2}}$. Alice and Bob
each receive the classical description of a POVM they have to perform on their respective qubit.
These can be represented by collections of vectors
$\mathcal{A}=\{\vec{a}_i|i=1,\ldots,r_A\}$ and $\mathcal{B}=\{\vec{b}_j|i=1,\ldots,r_B\}$ satisfying
conditions (\ref{eq:cond-povm1}-\ref{eq:cond-povm2}).
They then obtain measurement outcomes $A_{\textrm{pov}}\in\{1,\ldots,i,\ldots,r_A\}$
and $B_{\textrm{pov}}\in\{1,\ldots,j,\ldots,r_B\}$ respectively.
\end{defn}
According to quantum mechanics, the outcomes of Alice's and Bob's measurements,
$A_{\textrm{pov}}$ and $B_{\textrm{pov}}$, have the following joint probabilities:
\beqa
p(A_{\textrm{pov}}=i,B_{\textrm{pov}}=j|\mathcal{A},\mathcal{B})&=&\frac{|\vec{a_i}| |\vec{b_j}|- \vec{a_i}\cdot\vec{b_j}}{4},
\eeqa
with marginal probabilities \beqa
p(A_{\textrm{pov}}=i|\mathcal{A},\mathcal{B})&=&\frac{|\vec{a_i}|}{2},\\
p(B_{\textrm{pov}}=j|\mathcal{A},\mathcal{B})&=&\frac{|\vec{b_j}|}{2}.\\
\eeqa

We wish to simulate these correlations between $A_{\textrm{pov}}$ and $B_{\textrm{pov}}$ using a local hidden variable model
with the help of additional
resources. We have seen above that it is possible to simulate projective measurements on the singlet by sampling
or $f$-sampling a random vector
$\vec{\lambda_s}$ according to a biased distribution $\rho_{\vec{a}}(\vec{\lambda_s})={|\vec{a}\cdot\vec{\lambda_s}|}/{2\pi}$
 (Theorems~\ref{thm:sampling}
and~\ref{thm:strong-sampling}), that is,
if Alice and Bob have as input unit vectors $\vec{a_i'}=\vec{a_i}/|\vec{a_i}|$ and
$\vec{b_j'}=\vec{b_j}/|\vec{b_j}|$, and an infinite amount of shared randomness, they are able to produce,
with additional resources, random variables $A=\pm 1$
and $B=\pm 1$ with a joint distribution $p(A,B|\vec{a_i'},\vec{b_j'})= (1-AB\ \vec{a_i'}\cdot\vec{b_j'})/4$.
In order to simulate the POVMs, we will use one of the previous protocols
to simulate projective measures, then we will
test whether the outcomes of the projective measure protocol
``agree" with the POVM outcome we expected.   We call this test the POVM test.
In addition to the resources used for the
projective measure protocol, the POVM test will require further resources:
post-selection or communication. In this approach,
the test cannot be performed using only non-local boxes, because of their non-signaling property.
\begin{thm}
\label{thm:POVM}
It is possible for Alice and Bob to simulate the EPR experiment for POVMs
 using any one of  the following ressources.

\begin{itemize}
\item[(a)] postselection, with $p(\bot_A)=p(\bot_B)=2/3$;
\item[(b)] 6 bits of communication on average;
\item[(c)] 2 nl-bits plus 4 bits of communication, on average.
\end{itemize}
\end{thm}
\begin{proof}
Consider the following protocol.
\protocol{POVM protocol}{
Set $k=0$
\begin{enumerate}
\item Alice and Bob share a pair of uniformly distributed variables $\vec{\lambda_{k}},\vec{\lambda_{k+1}} \sim \sqcup_{\mathbb{S}_{2}}$,
\item Alice picks $\vec{a}_i\in\mathcal{A}$ following the marginal probabilities $p(i)=|\vec{a_i}|/2$.
\item Bob picks $\vec{b}_j\in\mathcal{B}$ following the marginal probabilities $p(j)=|\vec{b_j}|/2 $.
\item Alice and Bob perform a protocol to simulate the projective measurements on a singlet
with additional resources (either post-selection, communication or a non-local-box)
using $\vec{a_i'}=\vec{a_i}/|\vec{a_i}|$ and $\vec{b_j'}=\vec{b_j}/|\vec{b_j}|$ as inputs,
and obtaining $A$ and $B$ as outputs,
\item POVM test:
  \begin{itemize}
  \item with post-selection:
    \begin{enumerate}
    \item If $A=1$, Alice outputs $A_{\textrm{pov}}=i$, otherwise she aborts,
    \item If $B=1$, Bob outputs $B_{\textrm{pov}}=j$, otherwise he aborts.
    \end{enumerate}
  \item with communication:
    \begin{enumerate}
    \item Alice  sends $A$ to Bob,
    \item Bob sends $B$ to Alice.
    \item Alice and Bob test whether $A=B$,
      \begin{itemize}
      \item if so, they output $A_{\textrm{pov}}=i$, and $B_{\textrm{pov}}=j$ respectively,
      \item otherwise, they return to step 1 with $k = k+2$.
      \end{itemize}
    \end{enumerate}
  \end{itemize}
\end{enumerate}}
Intuitively, Alice and Bob choose POVM outcomes $\vec{a_i}\in\mathcal{A}$ and $\vec{b_j}\in\mathcal{B}$
with the right marginal distribution, and they use the projective measurement protocol to
realize this outcome.
To do this, they  simulate the corresponding (normalized) projective measurements
$\vec{a_i'}$ and $\vec{b_j'}$ on the singlet. Alice (and similarly for Bob) then gets an output
$A(\vec{a_i'},\vec{\lambda_s})=\pm 1$, corresponding to a spin either parallel or antiparallel
to $\vec{a_i}$. While an outcome $+1$ corresponds to a valid POVM element $\vec{a_i}\in\mathcal{A}$,
$-1$ does not correspond to a POVM element, $-\vec{a_i}\notin\mathcal{A}$, so
{\em a priori}, Alice should only accept $+1$ outcomes.
In the POVM test, they test whether the outcomes of the corresponding projective measurements
represent valid POVM elements $\vec{a_i'}$ and $\vec{b_j'}$ (POVM test),
and if not, either start over again or abort if they use post-selection.

Finally, depending on the resource used by Alice and Bob to perform the projective measurement simulation,
we have the following results.

\noindent{\bf Post-selection}
If Alice and Bob use post-selection, using Theorem~\ref{thm:post-selection}
they may simulate the projective measure on a singlet with a symmetric and apparently independent
abortion probability of $1/3$ for both parties. Moreover, both Alice and Bob will abort
when $A$ or $B$ equals $-1$, which happens with probability $1/2$, so that they output only when the
POVM test is satisfied. Altogether, they finally use post selection with probability $2/3$,
$p(\bot_A)=p(\bot_B)=2/3$.  This result is new.

\noindent{\bf Communication}
Recall that an outcome $+1$ of the projective measurement
corresponds to a valid POVM element $\vec{a_i}\in\mathcal{A}$
However, considering both Alice and Bob's outcomes
together, we see that $A=B=-1$ happen with the same probability that $A=B=+1$, so that in
the end, Alice and Bob only have to reject their outcomes when $A\neq B$.
If Alice and Bob use communication,
from Theorem~\ref{thm:communication} we see that they must use one bit of
communication to simulate the projective measure on the singlet state, and two bits of communication
to perform the POVM test. As the POVM test has an average probability of being satisfied of $1/2$,
they must repeat the projective measurement protocol twice on average to satisfy the test. So,
they use six bits of expected communication. This is a rewriting of Methot's result~\cite{methot04}.

\noindent{\bf Non-local boxes and communication}
Considering the last protocol, we may wonder whether the six bits of expected communication may
be replaced by uses of a non-local box. Obviously, we may simulate the
projective measure on the singlet with $1$ nl-bit.
However, to perform the POVM test in our protocol,
two-way communication is necessary since to know whether
her outcome $A$ agrees with Bob's, Alice needs to know $B$ and therefore acquire information from Bob,
and vice versa. Hence, our POVM test cannot be done with non-local boxes only since they do not
allow signaling. In this scenario, the best we can achieve is a protocol that uses two nl-bits and four bits of communication on average. This result is new.
\end{proof}


\section{Conclusion}
In summary, we have shown that the problem of simulating quantum correlations
exhibited by the singlet state
using different resources reduces to a distributed sampling problem, more precisely, to an
$f$-sampling problem. We have seen that to perform distributed sampling from a biased
distribution $\rho_{\vec{a}}$ depending only on Alice's input,
Alice can first sample the biased distribution locally and then use an additional resource to share
the biased distribution with Bob. To locally sample
from the biased distribution $\rho_{\vec{a}}$, we have given a new method called the ``choice method".
Once the local sampling is done, Alice uses an additional
resource to share her sample with Bob in a simple manner. This approach allows us to develop a simple and unified view of the problem and therefore a better understanding of the role of the
different resources used to gauge non-locality. The distribution $\rho_{\vec{a}}$ has a crucial role in our approach and we may ask whether it is
the unique distribution that arises naturally from this problem.
We have seen that simulating the EPR experiment is equivalent to
$f$-sampling from $\rho_{\vec{a}}$, but is not equivalent to distributed sampling
from $\rho_{\vec{a}}$.  An intermediate sampling problem is for Alice to sample
from $\rho_{\vec{a}}$ and Bob to learn only $\textrm{sgn}(\vec{b}\cdot\vec{\lambda_s)}$.
It would be interesting to know whether this problem is equivalent to
simulating the EPR experiment or not.
Such a result would help
us understand other properties of this problem, such as the optimality of the protocols.
The POVM case also fits into our approach, but the protocol we have derived
is probably not optimal because it simply consists in adapting the projective measurement
protocol to the POVM case. It might be the case that protocols can be based on the $f$-sampling
of another biased distribution designed especially for POVMs.

Finally, this approach could also
help study more general cases of the problem, for instance for higher dimensional
states (such as maximally entangled qudit pairs), multipartite entangled states, or
partially entangled states.

\begin{acknowledgements}
The research was supported by the EU 5th framework program RESQ
IST-2001-37559, and by the ACI CR 2002-40 and ACI SI 2003-24 grants of
the French Research Ministry.
J.R. acknowledges support from the French INRIA.
\end{acknowledgements}

\bibliography{bibliographie}

\end{document}

%% file: epr.tex
\setlength{\unitlength}{0.4mm}
\begin{picture}(110,45)(0,10)
\thinlines

\put(15,20){\rnode{Alice}{\framebox(25,15){Alice}}}
\put(25,35){\rnode{Alice2}{\makebox(0,0){}}}
\put(30,35){\rnode{Alice3}{\makebox(0,0){}}}
\put(70,20){\rnode{Bob}{\framebox(25,15){Bob}}}
\put(80,35){\rnode{Bob2}{\makebox(0,0){}}}
\put(85,35){\rnode{Bob3}{\makebox(0,0){}}}
\put(0,45){\rnode{a}{\makebox(10,10){$\vec{a}$}}}
\put(100,45){\rnode{b}{\makebox(10,10){$\vec{b}$}}}
\put(50,45){\rnode{psi}{\makebox(15,15){$\ket{\psi^-}$}}}
\put(22.5,0){\rnode{A}{\makebox(10,10){$A$}}}
\put(77.5,0){\rnode{B}{\makebox(10,10){$B$}}}

\nccurve[arrowsize=4pt,arrowinset=0.6,angleA=0,angleB=90]{->}{a}{Alice2}
\nccurve[arrowsize=4pt,arrowinset=0.6,angleA=180,angleB=90]{->}{b}{Bob3}
\nccurve[arrowsize=4pt,arrowinset=0.6,angleA=180,angleB=90]{->}{psi}{Alice3}
\nccurve[arrowsize=4pt,arrowinset=0.6,angleA=0,angleB=90]{->}{psi}{Bob2}
\nccurve[arrowsize=4pt,arrowinset=0.6,angleA=270,angleB=90]{->}{Alice}{A}
\nccurve[arrowsize=4pt,arrowinset=0.6,angleA=270,angleB=90]{->}{Bob}{B}
\end{picture}

%% file: lhv.tex
\setlength{\unitlength}{0.4mm}
\begin{picture}(110,45)(0,10)
\thinlines

\put(15,20){\rnode{Alice}{\framebox(25,15){Alice}}}
\put(25,35){\rnode{Alice2}{\makebox(0,0){}}}
\put(30,35){\rnode{Alice3}{\makebox(0,0){}}}
\put(70,20){\rnode{Bob}{\framebox(25,15){Bob}}}
\put(80,35){\rnode{Bob2}{\makebox(0,0){}}}
\put(85,35){\rnode{Bob3}{\makebox(0,0){}}}
\put(0,45){\rnode{a}{\makebox(10,10){$\overrightarrow{a}$}}}
\put(100,45){\rnode{b}{\makebox(10,10){$\overrightarrow{b}$}}}
\put(50,45){\rnode{lambda}{\makebox(10,10){$\lambda$}}}

\put(22.5,0){\rnode{A}{\makebox(10,10){$A(\vec{a},\lambda)$}}}
 \put(77.5,0){\rnode{B}{\makebox(10,10){$B(\vec{b},\lambda)$}}}
\nccurve[arrowsize=4pt,arrowinset=0.6,angleA=0,angleB=90]{->}{a}{Alice2}
\nccurve[arrowsize=4pt,arrowinset=0.6,angleA=180,angleB=90]{->}{b}{Bob3}
\nccurve[arrowsize=4pt,arrowinset=0.6,angleA=180,angleB=90]{->}{lambda}{Alice3}
\nccurve[arrowsize=4pt,arrowinset=0.6,angleA=0,angleB=90]{->}{lambda}{Bob2}
\nccurve[arrowsize=4pt,arrowinset=0.6,angleA=270,angleB=90]{->}{Alice}{A}
\nccurve[arrowsize=4pt,arrowinset=0.6,angleA=270,angleB=90]{->}{Bob}{B}
\end{picture}